# Evolution of Interfacial Hydration Structure Induced by Ion Condensation and Correlation Effects

Han Li[1,2,3#], Zhi Xu[1,3#], Jiacheng Li[1,3], Alessandro Siria[4], Ming Ma[1,2,3*]

[1] Department of Mechanical Engineering, State Key Laboratory of Tribology in Advanced Equipment (SKLT), Tsinghua University, Beijing 100084, China

[2] Institute of Superlubricity Technology, Research Institute of Tsinghua University in Shenzhen, Shenzhen 518118, China

[3] Center for Nano and Micro Mechanics, Tsinghua University, Beijing 100084, China

[4] Laboratoire de Physique de l'Ecole normale Supérieure, ENS, Université PSL, CNRS, Sorbonne Université, Université de Paris, Paris, France.

[#]Han Li and Zhi Xu contributed equally.

**Abstract:** Interfacial hydration structures are crucial in wide-ranging applications, including battery, colloid, lubrication etc. Multivalent ions like $Mg^{2+}$ and $La^{3+}$ show irreplaceable roles in these applications, which are hypothesized due to their unique interfacial hydration structures. However, this hypothesis lacks experimental supports. Here, using three-dimensional atomic force microscopy (3D-AFM), we provide the first observation for their interfacial hydration structures with molecular resolution. We observed the evolution of layered hydration structures at $La(NO_3)_3$ solution-mica interfaces with concentration. As concentration increases from 25 mM to 2 M, the layer number varies from 2 to 1 and back to 2, and the interlayer thickness rises from 0.25 to 0.34 nm, with hydration force increasing from 0.27±0.07 to 1.04±0.24 nN. Theory and molecular simulation reveal that multivalence induces concentration-dependent ion condensation and correlation effects, resulting in compositional and structural evolution within interfacial hydration structures. Additional experiments with $MgCl_2$-mica, $La(NO_3)_3$-graphite and $Al(NO_3)_3$-mica interfaces together with literature comparison confirm the universality of this mechanism for both multivalent and monovalent ions. New factors affecting interfacial hydration structures are revealed, including concentration and solvent dielectric constant. This insight provides guidance for designing interfacial hydration structures to optimize solid-liquid-interphase for battery life extension, modulate colloid stability and develop efficient lubricants.

## Introduction

At solid-liquid interface for aqueous solution, induced by surface topography and intermolecular interaction, the liquid molecular arrangement may be structured into non-continuous quasi-discrete layers with total thickness of several nanometers, known as hydration structures[1-3]. As a universal physical phenomenon, the interfacial hydration structures directly affect the intermolecular, inter-particle and inter-surface interactions at solid-liquid interface[1]. Such effect manifests as a short-range oscillatory interaction, termed as hydration force[1], which is considered as a fundamental interaction at solid-liquid interface.

The hydration structure and force affect the energy exchanges at liquid-solid interface[4-14] and are crucial for various applications, with multivalent ions offering unique benefits over monovalent ions. For examples, in batteries[11-12, 15], the near-electrode hydration structures determines the cycle life and capacity[9-10], known as solid-electrolyte-interphase (SEI) [11-12, 15-16], where multivalent ions like $Mg^{2+}$, $Zn^{2+}$, $Al^{3+}$ exhibit higher energy density than monovalent $Li^+$ and $Na^+$ ions[12]. In colloid chemistry, the hydration force determines the energy barrier of colloid aggression, and multivalent ions like $Mg^{2+}$ or $La^{3+}$ tune the colloid stability more effectively than monovalent ions[17-19]. For mechanical engineering, hydration structures by $Al^{3+}$, $Cr^{3+}$, $Mg^{2+}$, $La^{3+}$ have been speculated as effective hydration lubricating films with lower friction coefficients and higher load capacities compared to monovalent ions, reported as hydrated super-lubricity[13, 20]. Additionally, in nanofluidic devices[4], surficants[6], materials growth[5], catalysis[10, 21-22] and other fields, multivalent ions also improve the solid-liquid interface performance. A common hypothesis exists that the unique hydration structures with multivalent ions contribute to the improvement.

However, such hypothesis lacks experimental supports. Until now, the measurement of interfacial hydration structures for aqueous solution with multivalent ions remains unexplored. Common experimental methods of interfacial liquids are spectroscopy (e.g. sum-frequency generation[23-25]) and squeeze-out flow methods (e.g. surface force appratus[26] and traditional atomic force microscope[27-29]).

Spectroscopy is sensitive to molecular composition and orientation, but exhibit low horizontal spatial resolutions of ~ 100 μm due to the limitation of light-spot area ($10^4$~$10^5$ μm$^2$ area [25]). Squeeze-out methods capture in situ structural and additional mechanical information along one dimension perpendicular to the substrate, but yield controversial results due to large errors caused by liquid fluctuations[27-29]. These limitations bring challenges for early hydration structure measurement. Compared to existing methods, three-dimensional atomic force microscopy (3D-AFM)[30-34], a recently-developed approach, can provide both in situ structural and mechanical information of interfacial hydration structures with molecular resolution (spatial resolution < 0.1 nm, mechanical resolution ~0.01 nN)[30-34]. While experiments on solutions with monovalent ions (like NaCl, KCl) have been reported[30, 32-35], studies on multivalent ions are still lacking and await explorations.

Here, we image the three-dimensional hydration structures of lanthanum nitrate (La(NO$_3$)$_3$) aqueous solutions on mica with molecular resolution using 3D-AFM. Our observations reveal that both the interlayer thickness and hydration force of hydration structure remarkably increase with concentration. These findings are effectively elucidated through theoretical model and molecular dynamics (MD) simulations. Interfacial ion condensation and correlation effects are identified as origins of our observations. Our conclusion drawn based on La(NO$_3$)$_3$-mica is also substantiated in La(NO$_3$)$_3$-graphite, Al(NO$_3$)$_3$-mica and MgCl$_2$-mica interfaces, validating the universality of our results.

## Results & Discussion

### Experimental results

The images of 3D-AFM for La(NO$_3$)$_3$-mica interface are drawn in Fig. 1. The concentrations of La(NO$_3$)$_3$ aqueous solutions are 25 mM, 100 mM, 250 mM, 2M, respectively. From Fig. 1b~e, there are layered structures parallel to the surface (darker stripes), representing the hydration structures. As concentration increases, the hydration structure varies. Figure 1f~i depict 2D sectional images along *y* axis at different *x* locations, reflecting the evolution of hydration structures more clearly. For 25 mM La(NO$_3$)$_3$ solution, two layers are observed in the hydration structure, with an interlayer thickness of 0.25 nm estimated with phase data, as shown in Fig. 1 and Fig. S2 in supplementary information (SI). For 100 mM solution, it reduces to only one layer. For 250 mM solution, some locations exhibit one layer, while the rest have two layers, with an interlayer thickness of 0.34 nm. Lastly, for 2 M solution, all areas are double-layered, with an interlayer thickness of 0.34 nm.

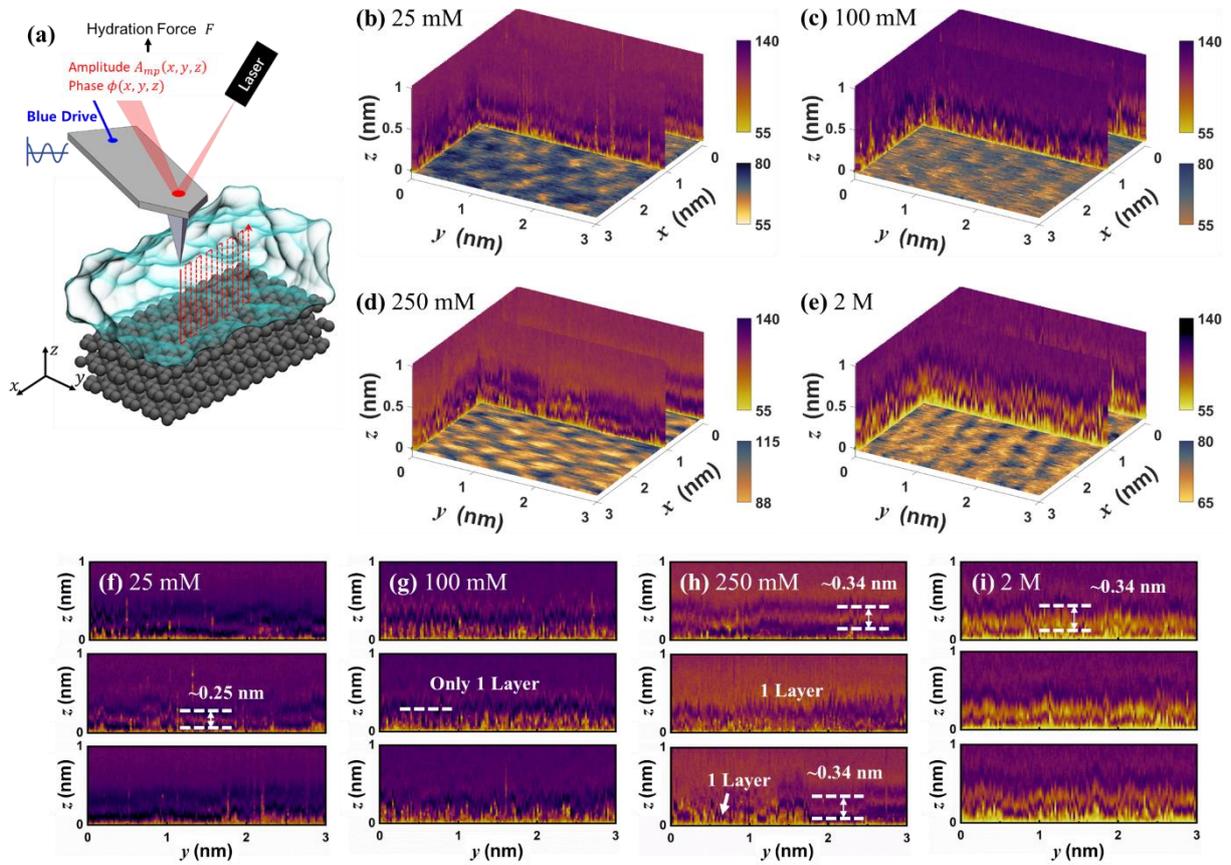

**Fig. 1: Schematic and results of 3D-AFM.** (a) Schematic of 3D-AFM measurement. (b~e) 3D image of layered hydration structures (darker stripes) for La(NO$_3$)$_3$-mica interface in 25 mM, 100 mM, 250 mM and 2M concentrations, respectively. (f~i) 2D sectional images of 25 mM, 100 mM, 250 mM and 2 M La(NO$_3$)$_3$-mica interfacial hydration structure at different locations, respectively.

Moreover, based on the 3D-AFM images, the hydration force $F$, which is the force experienced by the tip at a given distance $D$ away from the mica surface, can be computed and provides the quantitative mechanical feature of hydration structures. The calculation of $F$ follows Hölschera reconstruction method [31, 36], with details given in SI (section 1). To increase the signal-to-noise ratio of hydration force, the measured observables (amplitude, phase) are averaged for different $y$ positions at the same $z$[37]. Representative force-distance curves for different concentrations are given in Fig. 2a~d, showing notable oscillatory feature. The maximum peaks of hydration force (arrows) represent the locations and strengths of layered hydration structures, which can be fitted by Gaussian distribution, as shown in the dashed lines. As concentration increases, the distance between peaks increases from 0.25 nm

(Fig. 2a) to 0.34 ~ 0.4 nm (Fig. 2b~d), which is consistent with the observation in Fig. 1. The 2D *xz* maps of hydration force *F* are plotted in Fig. 2e~h, each comprising 64 force-distance curves at different *x* locations. From the 2D maps, the statistics of peak-value of hydration forces are calculated and observed to increase from 0.27 ± 0.07 nN (25 mM) to 1.04 ± 0.24 nN (2 M), as shown in Fig. 3a (blue solid line). To explain our findings, further analysis is proposed in the following discussions.

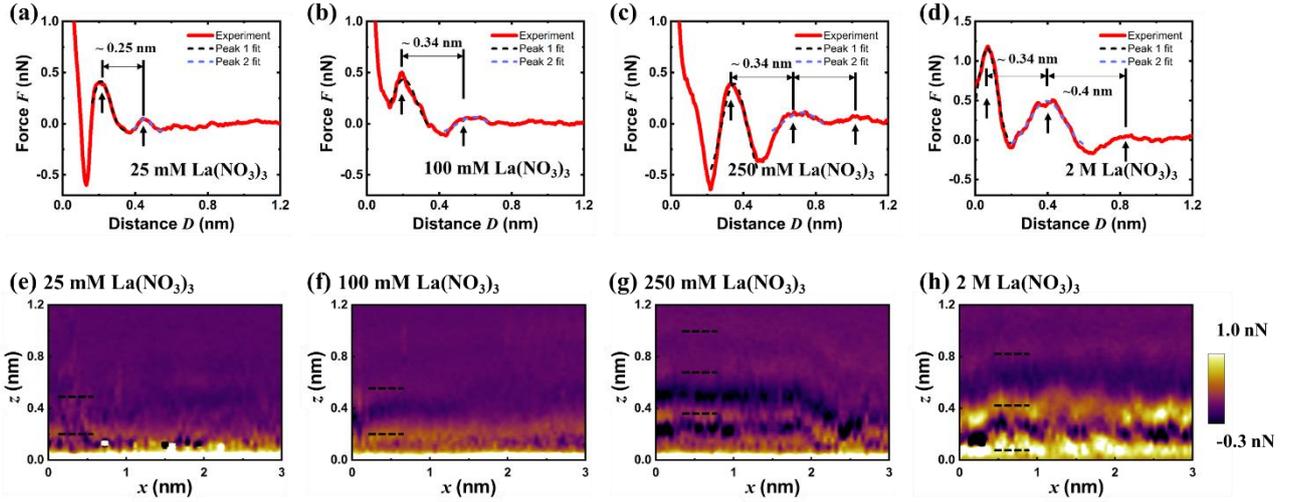

**Fig. 2: Hydration force of La(NO$_3$)$_3$-mica interface.** (a~d) Typical hydration force-distance curves in solutions with different concentration, the arrows represent the peaks of hydration forces, and the dashed lines represent the Gaussian fit of peaks. (e~h) 2D *xz* maps of hydration force. The dashed lines in (e~h) represent the location of layers, corresponding to the locations of peaks in (a~d) respectively.

**Mechanism of hydration force increasement: ion condensation**

Theoretical comprehension of the variation of hydration structures has been a longstanding challenge[1], no universal theory exists at present. To explain our observation, we begin from the charge adsorption mechanism of electrolyte solutions[1]. Due to the nanoscale morphology of AFM tip, the hydration on tip is much weaker than on mica[32]. Therefore, the mica substrate can be considered as an isolated surface. Based on the Poisson-Boltzmann-Grahame framework[1, 38], the surface charge density $\sigma$ of mica in La(NO$_3$)$_3$ solutions can be estimated as (detailed deviation is given in SI section 2):

$$\sigma^2 = 2\varepsilon\varepsilon_0 k_B T \rho_{b,La^{3+}} \left[ \exp\left(\frac{-3e\psi_0}{k_B T}\right) + 3\exp\left(\frac{e\psi_0}{k_B T}\right) - 4 \right] \quad (1)$$

where $\varepsilon, \varepsilon_0$ are the relative and vacuum dielectric constant, $k_B$ is Boltzmann constant, $T$ is temperature,

$\psi_0$ is surface potential, $\rho_{b,La^{3+}}$ is the bulk number density of $La^{3+}$ ions. The relative dielectric constant $\varepsilon$ is calculated by R. Adar model[39], and the surface potential $\psi_0$ is estimated by R. Pashley's work (details are shown in SI, section 2)[40]. Finally, $\sigma$ can be computed by equation (1).

Based on equation (1), as concentration increases, the surface charge density values are found to be positive and monotonically increase from +0.069 C/m² (25 mM) to +0.298 C/m² (2 M) (Fig. 3a, red dashed-dotted line), with corresponding increasement of surface number density for $La^{3+}$ ions from 1.55 nm$^{-2}$ to 2.03 nm$^{-2}$ (SI section 2.5). These results indicate that the $La^{3+}$ may condense at interface as concentration increases, known as ion condensation effect [41]. This effect can lead to non-repulsive electrical double layer force[1, 41-42], which is confirmed by additional DLVO force measurement as shown in Fig. S5 in SI.

From Fig. 3a, both hydration force and surface charge density increase monotonically with concentration $c$. According to contact value theory[1], the pressure between two surfaces immersed in electrolyte solutions can be estimated as: $P(D) = k_B T[\rho_0(D) - \rho_0(\infty)]$, where $D$ is separation between surfaces, $\rho_0(D)$ is the number density of ions at the middle position between two surfaces, $\rho_0(\infty)$ is equal to bulk ion concentration $\rho_b$. In our experiments, the first peak of hydration force are observed at small separation ($D < 0.1$ nm), where the number density of ions can be regarded as constant and denoted as $\rho_s$. For an isolated surface, based on Gauss's Law[1], $\rho_s = \rho_b + \frac{\sigma^2}{2\varepsilon\varepsilon_0 k_B T}$. Therefore, the hydration force at peak locations can be estimated:

$$F(D) = A k_B T (\rho_s - \rho_b) = \frac{A\sigma^2}{2\varepsilon\varepsilon_0} \qquad (2)$$

Here, the surface charge density $\sigma$ is calculated from equation (1), $A$ is effective area of AFM tip. With a fitted value of 13.5 nm² (3.67×3.67 nm²) for $A$, the outcomes of equation (2) semi-quantitatively agrees with experimental values, as shown in Fig. 3a (blue dashed line). Given that the tip radius is <10 nm, such fitting of $A$ is acceptable. Such consistency affirms the rationality of our analysis. In conclusion, the hydration force increasement is due to the rise of interfacial surface charge density, stemming from remarkable ion condensation effect in multivalent solutions.

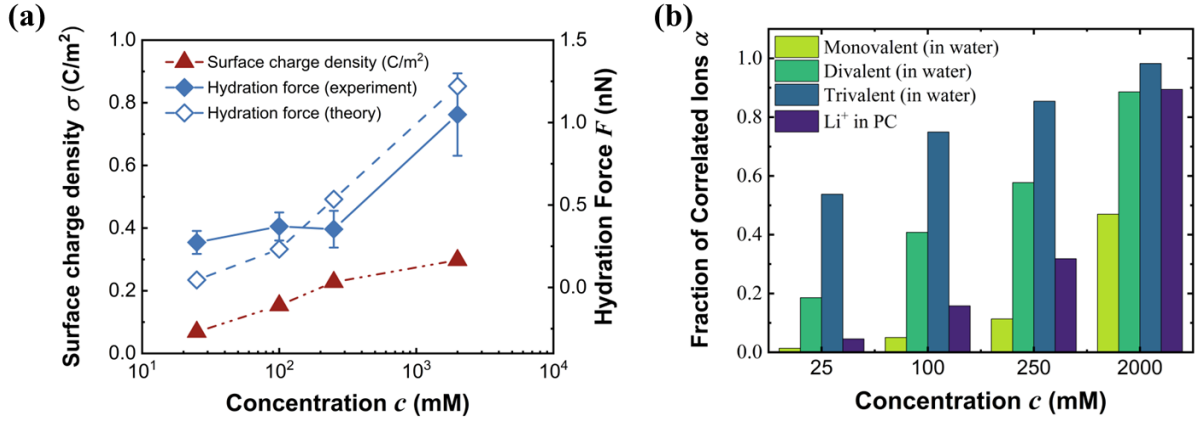

**Fig. 3: Mechanisms of hydration structures evolution.** (a) Ion condensation: Surface charge density $\sigma$ and hydration force $F$ variation calculated by equation (1 and 2), the effective surface area $A$ is fitted as 13.5 nm$^2$ (3.67×3.67 nm$^2$). (b) Ion correlation: Fraction of correlated ions $\alpha$ in bulk solutions with different ion-valences and solvents which are calculated using the methods proposed by Bjerrum[43] and Fuoss[44-45]. The theoretical results for monovalent ions are applicable to existing references[9, 33-34].

**Mechanism of hydration structure evolution: ion correlation**

As concentration increases, the condensed multivalent cations may attract anions conversely, leading to the formation of ion correlations (also called ion pairs)[45-46]. The propensity of ion correlation in bulk liquids can be quantified by the fraction of correlated ions, denoted by $\alpha$. When $\alpha = 1$, it means all the ions are paired. Two calculation methods (Bjerrum[43] and Fuoss[44]) of $\alpha$ are applied here[45], with details given in SI (Fig. S4). Figure 3b depicts $\alpha$ values at solutions with different valences and solvents. In trivalent aqueous solutions, $\alpha$ rises from 53% to 98% with concentration ranges from 25 mM to 2 M. The concentration of ion correlation in bulk liquids can be estimated as: $c_{ic} = \alpha c$. For 25 mM La(NO$_3$)$_3$ solution, $\alpha = 53\%$, $c_{ic} = 13.4$ mM, dissociated or partial-hydrated La$^{3+}$ ions are main components in solute. For 2 M La(NO$_3$)$_3$ solution, $\alpha = 98\%$, $c_{ic} = 1.96$ M, the La$^{3+}$-(NO$_3^-$) correlations become dominant in solute, potentially thickening the hydration structure.

The Bjerrum and Fuoss methods only account for bulk liquids[45]. Considering the differences of molecular arrangement between interfacial and bulk liquids, molecular dynamic (MD) simulations are conducted. The MD models are depicted in Fig. 4a, with details provided in SI (section 5). Ions and molecules within 1.2 nm from the mica surfaces are identified as interfacial particles according to experimental observations (Fig. 2a-d). As shown in Fig. 4, in the 25 mM solution, the interfacial region mainly comprises La$^{3+}$ and water molecules. As concentration increases, NO$_3^-$ ions becomes more pronounced and correlates with La$^{3+}$. Quantitatively, the effect of ion correlation can be described by radial distribution functions between interfacial particles, denoted as $g(r) = \langle\rho(r)\rangle/\rho$, with $\langle\rho(r)\rangle$ being local number density of particles at a distance $r$, and $\rho$ being the bulk density of particles. As shown in Fig. 4b, $g(r)$ between La$^{3+}$ ions and oxygen atoms of water molecules (La-Ow) exhibit notable sharp peaks at the location of $r = 0.26$ nm for all concentrations, representing the water shell around La$^{3+}$ ion. In comparison, $g(r)$ for La$^{3+}$ ions and N atoms in nitrate ions (La-N) vary with concentration (Fig. 4c). Solutions with concentrations of 100 mM to 2 M show sharp peaks at $r = 0.37$ nm, indicating the existence of interfacial La$^{3+}$-(NO$_3^-$) correlation. While for 25 mM solution, such peak disappears, suggesting the absence of La$^{3+}$-(NO$_3^-$) correlation at interface. As a result, the impact of La$^{3+}$-(NO$_3^-$) ion correlation on the hydration structure for 25 mM solution is weak, with an interlayer thickness close to La-Ow distance (0.25 nm in experiments vs 0.26 nm in MD). Conversely, the measured structure in 2 M solution is dominated by La$^{3+}$-(NO$_3^-$) correlation, with an interlayer thickness close to La-N distance (0.34 nm in experiments vs 0.37 nm in MD). The results of intermediate concentration are transition between two extremes.

The ion correlation can be denoted as La-(NO$_3$)$_m$, with $m$ being coordination number[47], which can be calculated based on $g(r)$. Here, $m$ are counted as 0.21, 0.62, 0.69, and 3.49 for concentration of 25 mM, 100 mM, 250 mM and 2 M respectively, indicating a clear increase with concentration. Known the ion quantity (SI section 5.2) and $m$, the concentration of La$^{3+}$-(NO$_3^-$) correlation within the interfacial region is estimated, i.e. the density of NO$_3^-$ ions which are paired, revealing a 35-times increase as concentration increases from 25 mM to 2 M (Fig. 4d). Due to interfacial ion condensation[48], the calculated interfacial correlation concentration is higher than bulk value ($c_{ic}$). The corresponding structures of interfacial ion correlations are drawn in the insets of Fig. 4d. Ultimately, the MD simulation results support the previous theoretical conclusion: the impact of ion correlation is weak for dilute 25 mM solution but remarkable for concentrated 2 M solution.

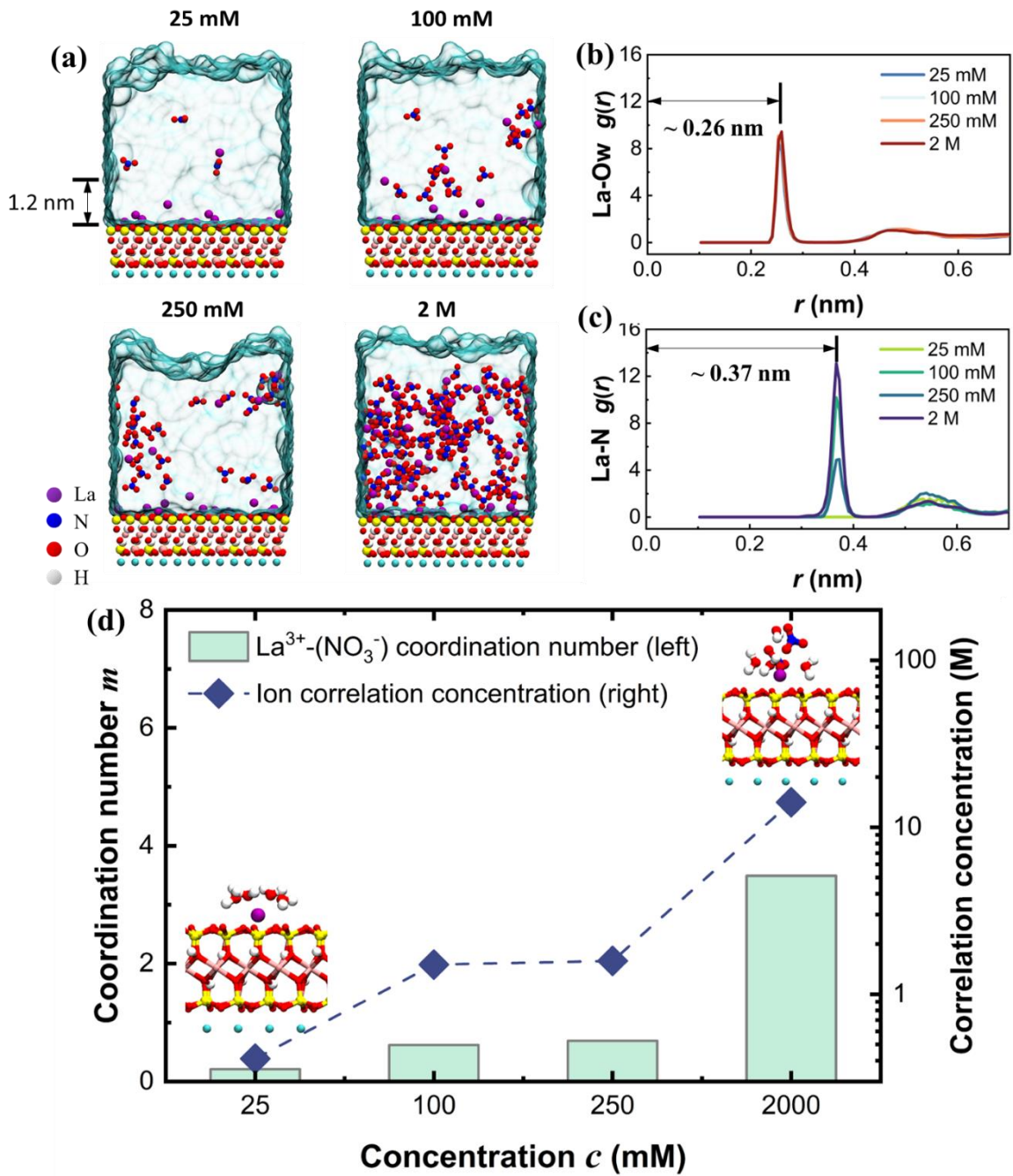

**Fig. 4: Molecular dynamic simulations for ion correlation effect.** (a) Models used in simulation for solutions with different concentration. (b) Radial distribution functions $g(r)$ between $La^{3+}$ ions and oxygen atoms of water molecules (La-Ow). (c) Radial distribution functions $g(r)$ between $La^{3+}$ ions and N atoms in nitrate ions (La-N). (d) Coordination number and concentration of $La^{3+}$-$(NO_3^-)$ correlation, insets reveal the typical structures of interfacial correlation.

The above discussion provides an inference that the composition of the two hydration layers varies with hydration structure evolution. In 25 mM solution, both layers primarily consist of $La^{3+}$ and water.

In the 2 M solution, the 1st layer (closest to the substrate) mainly consists of $La^{3+}$ ions and water, while the 2nd layer (further from the substrate) primarily consists of $NO_3^-$ ions and water. This inference is supported by further analysis. In Fig. 2a-d, the peaks of experimental results are fitted by Gaussian distribution, with peak width quantified by the full width at half maximums (FWHMs), as presented in Table 1. FWHMs for the density profiles of different ions and molecules in MD simulations (details in SI section 5.4) are also fitted for comparison. For 25 mM solution, experimental FWHMs of the two peaks are close to simulated FWHMs of $La^{3+}$ and water. For the 100 mM, 250 mM and 2 M solutions, experimental FWHMs of 1st peak fall between simulated FWHMs of $La^{3+}$ and water, while experimental FWHMs of 2nd peak increase and closely match simulated FWHMs of $NO_3^-$ and water, which is consistent with the inference based on interlayer distance and fraction of correlated ions. Consequently, together with the experimental, theoretical and simulation results, the ion correlation effect is clear: it alters the composition within the interfacial hydration structures, leading to structural evolution, which is experimentally observed as interlayer thickening.

**Table 1. Full width at half maximum comparison between experiments and simulations.**

| Concentration | FWHMs (experiments) | FWHMs (simulation) |
|---|---|---|
| 25 mM | 0.11 nm (1st peak) | 0.11 nm ($La^{3+}$) |
|  | 0.09 nm (2nd peak) | 0.17 nm (water) |
|  |  | 0.24 nm ($NO_3^-$) |
| 100 mM | 0.16 nm (1st peak) | 0.08 nm ($La^{3+}$) |
|  | 0.22 nm (2nd peak) | 0.22 nm (water) |
|  |  | 0.52 nm ($NO_3^-$) |
| 250 mM | 0.16 nm (1st peak) | 0.13 nm ($La^{3+}$) |
|  | 0.26 nm (2nd peak) | 0.22 nm (water) |
|  |  | 0.25 nm ($NO_3^-$) |
| 2 M | 0.12 nm (1st peak) | 0.05 nm ($La^{3+}$) |
|  | 0.24 nm (2nd peak) | 0.23 nm (water) |
|  |  | 0.27 nm ($NO_3^-$) |

**Comparison with existing results in monovalent ionic solutions**

Besides our observations, the proposed ion condensation and correlation mechanisms can further explain the divergent results on the concentration-dependent hydration structure of monovalent ions[9, 33-34]. For example, Z. Li et al found that the interlayer thickness of hydration structures of KCl, LiCl and NaCl aqueous solutions remained independent of concentration (details shown in SI)[33-34]. However, observations in LiTFSI-propylene carbonate (PC) solutions by Y. Yamagishi suggest remarkable interlayer thickening[9]. As depicted in Fig. 3b, the ion correlation mechanism reveals that the fraction of correlated ions $\alpha$ amounts to 1%, 5%, 11%, and 47% in 25 mM, 100 mM, 250 mM, and 2M monovalent aqueous solutions, respectively, which is significantly lower than in divalent and trivalent solutions. As a result, for the experiments by Z. Li et al, the ion correlation effect on interlayer thickness is secondary. In comparison, due to the lower solvent dielectric constant ($\varepsilon = 60$), monovalent ions in PC exhibits a high $\alpha$ for concentrated solutions (89% for 2 M), consistent with the observed interlayer thickening[9]. This consistency validates the rationality of the proposed ion correlation mechanism. Based on this mechanism, ionic valence, concentration, and solvent dielectric constant can be utilized to modulate the interfacial hydration structures. It is worth noting that besides the divergent interlayer thickness variation, the previously reported hydration forces all increase with concentration[9, 33-34], validating the ion condensation mechanism. In summary, compared with monovalent ions, the uniqueness of multivalent ions lies in the stronger ion correlation effect denoted by higher $\alpha$ and the resulting evolutionary interfacial hydration structure.

**Universality of ion condensation and correlation mechanisms**

To verify the universality of our findings, we perform additional experiments at different solid-liquid interfaces, including $MgCl_2$-mica, $La(NO_3)_3$-graphite and $Al(NO_3)_3$-mica, respectively. The results, as shown in Fig. 5, reveal similar trends. For 25 mM solutions, the interlayer thickness of hydration structures is measured as 0.16 nm for $MgCl_2$, 0.49 nm for $La(NO_3)_3$, and 0.36 nm for $Al(NO_3)_3$, with hydration force (peak-values) being 0.59 nN, 0.2 nN and 0.51 nN, respectively. Upon increasing the concentration to 2 M, the thickness values increased to 0.33, 0.57 and 0.46 nm, with hydration force (peak-values) increasing to 0.67, 0.49 and 0.69 nN, respectively. Both interlayers thickness and hydration force increase, which is consistent with the results measured at $La(NO_3)_3$-mica interfaces. The observed results across different types of solid-liquid interfaces align closely with our discussion about ion condensation and correlation, providing direct experimental support for the universality of our conclusion and reinforcing the robustness of our findings.

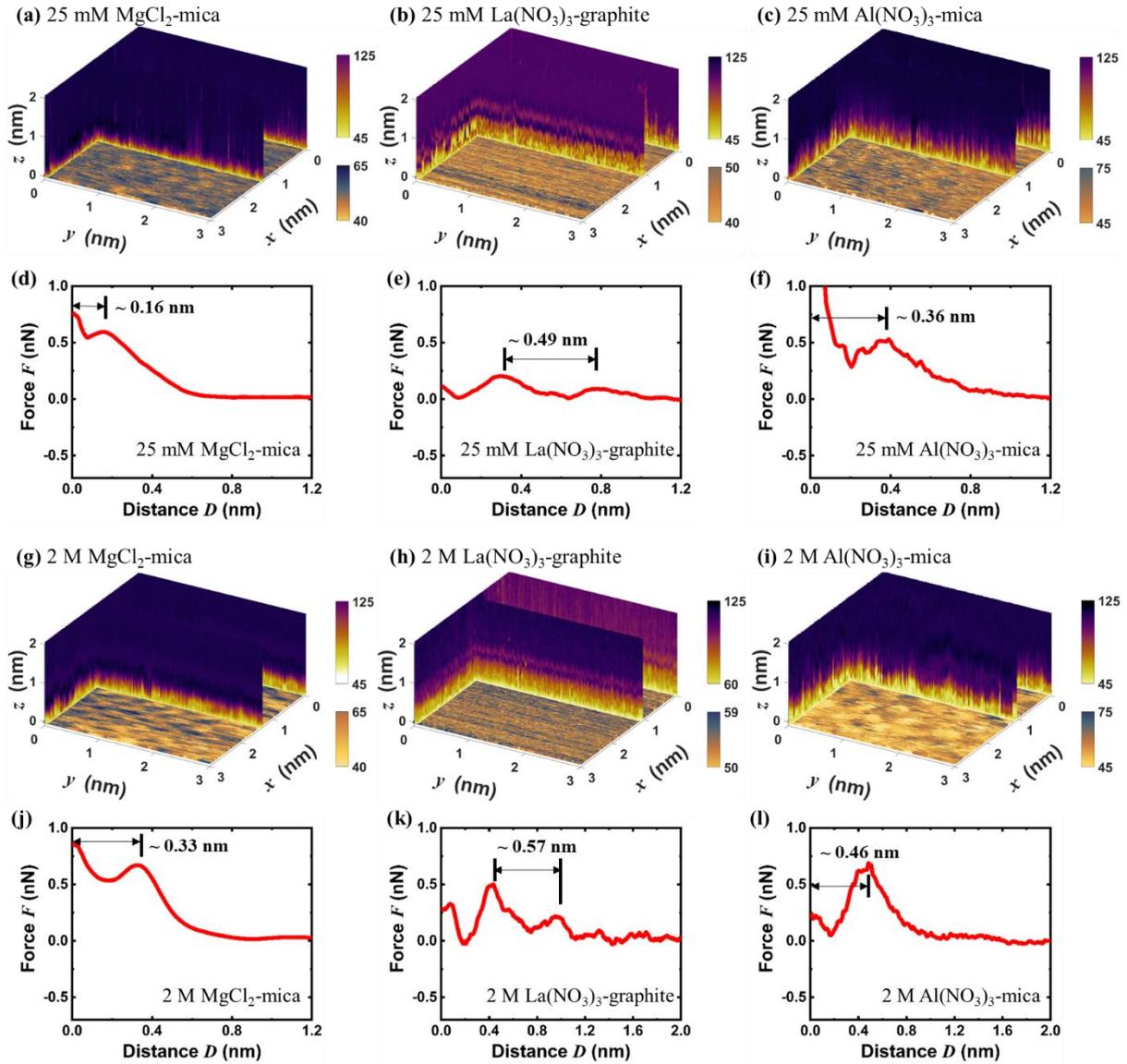

**Fig. 5: Hydration structures and forces at MgCl$_2$-mica, La(NO$_3$)$_3$-graphite, and Al(NO$_3$)$_3$-mica interfaces.** (a~c) 3D-AFM images of hydration structures in 25 mM solutions. (d~f) Hydration forces in 25 mM solutions. (g~i) 3D-AFM images of hydration structures in 2 M solutions. (j~l) Hydration forces in 2 M solutions. The interlayer thickness and hydration force increase with concentrations at all solid-liquid interfaces.

## Conclusion

In summary, we first experimentally investigate hydration structures in La(NO$_3$)$_3$ solution and observe an evolution of hydration structure, symbolized by the variation in number of layers and the interlayer thickness, resulting in the increasement of hydration force. Theory and molecular simulation analyses reveal that multivalence induces remarkable concentration-dependent ion condensation and correlation effects at interface, resulting in compositional and structural evolution within hydration structures, experimentally observed as hydration force increasement and interlayer thickening. These mechanisms not only explain our findings but also apply to monovalent ionic solutions. Additional experiments at three different solid-liquid interfaces ($MgCl_2$-mica, $La(NO_3)_3$-graphite and $Al(NO_3)_3$-mica) confirm the universality of our results.

This work provides the first observation of hydration structure in multivalent ionic solutions with molecular resolution, filling the absense and providing direct supports for the hypothesis that multivalent ions exhibit unique interfacial hydration structures. Compared with monovalent ions, the uniqueness of multivalent ions lies in the stronger ion correlation effect denoted by higher $\alpha$ and the resulting evolutionary interfacial hydration structure. This insight advances the understanding and promotes discovery for the modulating approaches of interfacial liquid structures using ion correlation effect, such as ionic valence, concentration, and solvent dielectric constant. Further studies can be guided to the design of functional hydration structures in various applications, such as optimizing SEI for battery life, designing the colloid stabilization and developing efficient lubricants.

## Methods

### Sample preparation

In experiments, deionized water (18.2 MΩ, Hitech Sciencetool) is used as solvent for solution preparation. Aqueous solutions with different concentrations and solutes are degassed under vacuum environments (0.01 atm, 25 °C) before AFM experiments. Muscovite mica (V-1 grade, Tedpella) and highly oriented pyrolytic graphite (HOPG, ZYB quality) are applied as solid substrates, respectively. Before measurement, the substrates are freshly cleaved to get a clean surface. Arrow-UHF AFM probe (NanoWorld) is used for AFM measurement, with a spring constant of 6 N/m, a tip radius <10 nm and a resonance frequency of 450 kHz in aqueous solutions.

### AFM experiments

3D-AFM images are measured in an amplitude-modulated atomic force microscope (Cypher ES

AFM, Oxford Instrument). During experiments, the AFM probe and measured substrate are immersed in given aqueous solutions. A blue laser is focused on the root of the probe cantilever to drive a harmonic vibration with a free amplitude $A_{mp0}$ of 0.3 nm (Blue Drive, Fig. 1a and Fig. S1a in SI). The setpoint of amplitude modulated AFM scanning $A_{mps}$ is set as $A_{mps}/A_{mp0} = 0.9$~$0.95$, which is the same as R. Garcia[37]. The piezoelectric ceramic scanner installed on the sample stage drives the substrate approaches to and then retracts from the probe. By scanning point-by-point with a $xy$ scan-rate of 39 Hz and $z$ scan-rate of 20 Hz, the amplitude and phase of cantilever are recorded by the optical lever system of AFM, providing 3D structural and mechanical information. Here, the $x$ and $y$ axes are defined as directions parallel to the solid surface, and the $z$ axis are perpendicular to surface. The $xy$ scanning area is 3×3 nm$^2$, and the $z$ scanning distance is 5 nm. The resolution of 3D-AFM image is 64×128×1600 for the rectangular measured space of 3×3×5 nm$^3$. Compared to amplitude data, phase data are more sensitive and therefore utilized for three-dimensional imaging[16, 30], as shown in Fig. 1. The phase shift represents the mechanical characteristics perceived by AFM tip, such as force and dissipation[36].

The raw data (phase and amplitude) of AFM can be further transformed as hydration force. According to the method proposed by Hölschera[31, 49], the hydration force $F$ between tip and substrate can be calculated as [31, 36, 49]: $F(D) = \frac{\partial}{\partial D} \int_D^{D+2A_{mp}} \frac{kA_{mp}^{1.5}\left[\frac{a_d \cos(\varphi)}{\sqrt{2}} - \frac{f_0^2 - f_d^2}{f_0^2}\right]}{\sqrt{z_D - D}} dz_D$, where $D$ is the nearest separation between the tip and the sample in one vibration period, $z_D$ is the average height of the probe in one vibration period, $A_{mp}$ is the amplitude of the probe, $\varphi$ is the phase of the probe, $a_d$ is the amplitude of the root of the probe, $k$ is the stiffness of the probe, $f_0$ is the natural frequency of the probe cantilever beam, and $f_d$ is the driving frequency. $A_{mp}$ and $\varphi$ are raw data measured by AFM, respectively. To minimize the effect of noise and improve signal-to-noise ratio, we applied the same method as M. Uhlig and R. Garcia al[37]. For each 3D-AFM image, it contains 64×128=8192 force-distance curves. We average the measured phase and data for each 128 force-distance curves, and get $\overline{A_{mp}}$ and $\overline{\varphi}$, with $\overline{A_{mp}}$ and $\overline{\varphi}$ being averaged amplitude and phase respectively. By substituting $\overline{A_m}$ and $\overline{\varphi}$ to the equation of $F$ we can get 64 hydration force curves for each 3D-AFM image, then we plot the 2D maps of hydration force, as shown in Fig. 2.

Besides 3D-AFM, additional DLVO measurements are conducted to confirm the existence of ion condensation. The DLVO measurements are based on colloid probe method, with a spherical probe which is composed of a tipless cantilever (CSC38, Mikromasch Ins.) and a microsphere (borosilicate glass, radius 10 μm, Duke scientific). More details of experiments are shown in SI (section 1 and section 3).

**MD simulations**

The MD model for La(NO$_3$)$_3$–mica system is shown in Fig. S6 in SI. Periodic boundary condition is applied to *x*, *y* and *z* directions with size of 27 × 31 × 210 Å$^3$. The water slab with height of ~3 nm is put on mica surface. A vacuum layer of 16 nm is applied to avoid long range coulomb force from periodic molecules in *z* direction. CLAYFF[50] and SPC/E[51] force fields are used for mica and water respectively, which had been applied widely to study the water-mica interactions. The Lennard-Jones potential is applied to La$^{3+}$ ions, with parameters[52] refined by directly comparing the hydration structure obtained from the simulations with the extended X-ray absorption fine structure experimental data. OPLS-AA force field is set for NO$_3^-$ ions[53], showing ability to predict properties of ammonium nitrate aqueous solutions. Lorentz–Berthelot mixing rules are used for the inter-molecular interactions that is not given explicitly. All the MD simulations are carried out with LAMMPS packages[54] using a time step of 1 fs. Truncated and force-shifted LJ interaction that combines the standard Lennard-Jones function and subtracts a linear term based on the cutoff distance is used here[55], so that both, the potential and the force, go continuously to zero at the cutoff distance of 1.2 nm. Long-range Columbic interactions are computed using the particle–particle particle–mesh (PPPM) algorithm.

Temperature of 298 K is maintained by applying the Nosé-Hoover chain thermostat to the liquid. Mica is considered as rigid during the simulations since its large stiffness. 5 ns simulations are conducted to collect interfacial structure information after 1 ns equilibration. According to the Fig. 3, surface charge density $\sigma$ can be calculated based on the Poisson-Boltzmann framework, as 0.07-, 0.15-, 0.23- and 0.30 C/m$^2$ for 25-, 100-, 250- and 2000 mM solution respectively. We put care to make sure that the surface charge density between experiments and simulations are consistent, since it will affect liquid structure significantly. First, La$^{3+}$ ions are put on mica surface randomly to result in a positive charged surface with the same $\sigma$ of experiments. Then ionic aqueous solution is added on the top of prepared surface. Lastly, additional NO$_3^-$ ions are used to guarantee electrical neutrality for the whole system, avoiding artificial long range Coulomb interaction between mirror molecules. More details of MD simulation are provided in SI section 5.

**Peak width analysis**

Classical analysis of experimental hydration force only focused on the hydration force magnitude and interlayer thickness. No studies perceived the variation of peak width. Here, in our experiments, we find that the peak width of hydration force varies with tip-sample separation and concentration.

The peak width is quantitatively described by full width at half maximum (FWHM). Here, the peak region in force-distance curve is defined as the region between two adjacent minimum points, in which the maximum value is defined as peak value of hydration force. The peak region can be fitted by a Gaussian normal distribution with a standard deviation of $\sigma_N$. For Gaussian distribution, FWHM = $2.355\sigma_N$. In Fig. 2a-d in the main-text, the peaks experimental force-distance curves indeed follow the Gaussian form, with FWHMs fitted as shown in Table 1 in the main-text. For comparison purpose, we also fit the FWHMs of three different particles ($La^{3+}$, $NO_3^-$, water) in MD simulation using the density profiles. The details of FWHM results are shown in Fig. S8 in SI.

# References

1. Israelachvili, J. N., *Intermolecular and Surface Forces, 3rd Edition*. Elsewier: 2011; p 1-674.
2. Umeda, K.; Kobayashi, K.; Minato, T.; Yamada, H., Atomic-scale three-dimensional local solvation structures of ionic liquids. *The Journal of Physical Chemistry Letters* **2020,** *11* (4), 1343-1348.
3. Schlesinger, I.; Sivan, U., Three-dimensional characterization of layers of condensed gas molecules forming universally on hydrophobic surfaces. *Journal of the American Chemical Society* **2018,** *140* (33), 10473-10481.
4. Robin, P.; Kavokine, N.; Bocquet, L., Modeling of emergent memory and voltage spiking in ionic transport through angstrom-scale slits. *Science* **2021,** *373* (6555), 687.
5. Dighe, A. V.; Singh, M. R., Solvent fluctuations in the solvation shell determine the activation barrier for crystal growth rates. *Proceedings of the National Academy of Sciences of the United States of America* **2019,** *116* (48), 23954.
6. Huang, K.; Rowe, P.; Chi, C.; Sreepal, V.; Bohn, T.; Zhou, K.-G.; Su, Y.; Prestat, E.; Pillai, P. B.; Cherian, C. T.; Michaelides, A.; Nair, R. R., Cation-controlled wetting properties of vermiculite membranes and its promise for fouling resistant oil–water separation. *Nature Communications* **2020,** *11*, 1097.
7. Cui, H. Y.; Zhang, L. L.; Eltoukhy, L.; Jiang, Q. J.; Korkunç, S. K.; Jaeger, K. E.; Schwaneberg, U.; Davari, M. D., Enzyme hydration determines resistance in organic cosolvents. *Acs Catalysis* **2020,** *10* (24), 14847-14856.
8. Abellán-Flos, M.; Tanç, M.; Supuran, C. T.; Vincent, S. P., Multimeric xanthates as carbonic anhydrase inhibitors. *Journal of Enzyme Inhibition and Medicinal Chemistry* **2016,** *31* (6), 946-952.
9. Yamagishi, Y.; Kominami, H.; Kobayashi, K.; Nomura, Y.; Igaki, E.; Yamada, H., Molecular-resolution imaging of interfacial solvation of electrolytes for lithium-ion batteries by frequency modulation atomic force microscopy. *Nano Letters* **2022,** *22*, 9907-9913.
10. Bonagiri, L. K. S.; Panse, K. S.; Zhou, S.; Wu, H. Y.; Aluru, N. R.; Zhang, Y. J., Real-space charge density profiling of electrode-electrolyte interfaces with angstrom depth resolution. *ACS Nano* **2022,** *16* (11), 19594-19604.
11. Rajput, N. N.; Seguin, T. J.; Wood, B. M.; Qu, X. H.; Persson, K. A., Elucidating solvation structures for rational design of multivalent electrolytes-a review. *Topics in Current Chemistry* **2018,** *376*, 19.
12. Li, S. Y.; Zhang, J. H.; Zhang, S. C.; Liu, Q. L.; Cheng, H.; Fan, L.; Zhang, W. D.; Wang, X. Y.; Wu, Q.; Lu, Y. Y., Cation replacement method enables high-performance electrolytes for multivalent metal batteries. *Nature Energy* **2024,** *9* (3), 285-297.
13. Han, T.; Cao, W.; Xu, Z.; Adibnia, V.; Olgiati, M.; Valtiner, M.; Zhang, C.; Ma, M.; Luo, J.; Banquy, X., Hydration layer structure modulates superlubrication by trivalent $La^{3+}$ electrolytes. *Science Advances* **2023,** *9* (28), 3902.
14. Zhang, Z.; Piatkowski, L.; Bakker, H. J.; Bonn, M., Ultrafast vibrational energy transfer at the water/air interface revealed by two-dimensional surface vibrational spectroscopy. *Nature Chemistry* **2011,** *3* (11), 888-893.
15. Suo, L.; Oh, D.; Lin, Y.; Zhuo, Z.; Borodin, O.; Gao, T.; Wang, F.; Kushima, A.; Wang, Z.; Kim, H.-C., How solid-electrolyte interphase forms in aqueous electrolytes. *Journal of the American Chemical Society* **2017,** *139* (51), 18670-18680.
16. Yamagishi, Y.; Ohuchi, S.; Igaki, E.; Kobayashi, K., Exploring the Molecular-Scale Structures at Solid/Liquid Interfaces of Li-Ion Battery Materials: A Force Spectroscopy Analysis with Sparse Modeling. *Nano Letters* **2024**.
17. Nikolov, A.; Lee, J. J.; Wasan, D., DLVO surface forces in liquid films and statistical mechanics of colloidal oscillatory structural forces in dispersion stability. *Advances in Colloid and Interface Science* **2023,** *313*.
18. Trefalt, G.; Szilagyi, I.; Tellez, G.; Borkovec, M., Colloidal stability in asymmetric electrolytes: Modifications of the Schulze–Hardy rule. *Langmuir* **2017,** *33* (7), 1695-1704.
19. Schneider, C.; Hanisch, M.; Wedel, B.; Jusufi, A.; Ballauff, M., Experimental study of electrostatically stabilized colloidal particles: Colloidal stability and charge reversal. *Journal of colloid and interface science* **2011,** *358* (1), 62-67.

20. Han, T.; Zhang, C.; Li, J.; Yuan, S.; Chen, X.; Zhang, J.; Luo, J., Origins of superlubricity promoted by hydrated multivalent ions. *The Journal of Physical Chemistry Letters* **2020,** *11* (1), 184-190.
21. Ringe, S.; Hörmann, N. G.; Oberhofer, H.; Reuter, K., Implicit solvation methods for catalysis at electrified interfaces. *Chemical Reviews* **2022,** *122*, 10777.
22. Andersson, M. P.; Stipp, S. L. S., Predicting hydration energies for multivalent ions. *Journal of Computational Chemistry* **2014,** *35* (28), 2070.
23. Ma, P.; Liu, Y.; Han, K.; Tian, Y.; Ma, L., Hydration lubrication modulated by water structure at $TiO_2$-aqueous interfaces. *Friction* **2023,** *12*, 591.
24. Litman, Y.; Chiang, K.-Y.; Seki, T.; Nagata, Y.; Bonn, M., Surface stratification determines the interfacial water structure of simple electrolyte solutions. *Nature Chemistry* **2024,** *16*, 644.
25. Shen, P. B. M. a. Y. R., Liquid Interfaces: A Study by Sum-Frequency Vibrational Spectroscopy. *Journal of Physical Chemistry B* **1999,** *103*, 3292.
26. Klein, J.; Kumacheva, E., Confinement-induced phase-transitions in simple liquids. *Science* **1995,** *269* (5225), 816-819.
27. Li, T.-D.; Gao, J.; Szoszkiewicz, R.; Landman, U.; Riedo, E., Structured and viscous water in subnanometer gaps. *Physical Review B* **2007,** *75* (11), 115415.
28. Goertz, M. P.; Houston, J. E.; Zhu, X. Y., Hydrophilicity and the viscosity of interfacial water. *Langmuir* **2007,** *23* (10), 5491-5497.
29. Jeffery, S.; Hoffmann, P. M.; Pethica, J. B.; Ramanujan, C.; Ozer, H. O.; Oral, A., Direct measurement of molecular stiffness and damping in confined water layers. *Physical Review B* **2004,** *70* (5).
30. Garcia, R., Interfacial liquid water on graphite, graphene, and 2d materials. *ACS Nano* **2023,** *17* (1), 51-69.
31. Fukuma, T.; Garcia, R., Atomic- and molecular-resolution mapping of solid-liquid interfaces by 3D atomic force microscopy. *ACS Nano* **2018,** *12* (12), 11785-11797.
32. Martin-Jimenez, D.; Chacon, E.; Tarazona, P.; Garcia, R., Atomically resolved three-dimensional structures of electrolyte aqueous solutions near a solid surface. *Nature Communications* **2016,** *7*, 12164.
33. Li, Z. B.; Liu, Q.; Li, Q.; Dong, M. D., The role of hydrated anions in hydration lubrication. *Nano Research* **2022,** *16*, 1096–1100.
34. Li, Z. B.; Liu, Q.; Zhang, D. L.; Wang, Y.; Zhang, Y. G.; Li, Q.; Dong, M. D., Probing the hydration friction of ionic interfaces at the atomic scale. *Nanoscale Horiz.* **2022,** *7* (4), 368.
35. Benaglia, S.; Uhlig, M. R.; Hernandez-Munoz, J.; Chacon, E.; Tarazona, P.; Garcia, R., Tip charge dependence of three-dimensional afm mapping of concentrated ionic solutions. *Physical Review Letters* **2021,** *127* (19), 196101.
36. Hoelscher, H., Quantitative measurement of tip-sample interactions in amplitude modulation atomic force microscopy. *Appl. Phys. Lett.* **2006,** *89* (12), 123109.
37. Uhlig, M. R.; Martin-Jimenez, D.; Garcia, R., Atomic-scale mapping of hydrophobic layers on graphene and few-layer $MoS_2$ and $WSe_2$ in water. *Nature Communications* **2019,** *10*, 2606.
38. Grahame, D. C., Diffuse double layer theory for electrolytes of unsymmetrical valence types. *The Journal of Chemical Physics* **1953,** *21* (6), 1054-1060.
39. Adar, R. M.; Markovich, T.; Levy, A.; Orland, H.; Andelman, D., Dielectric constant of ionic solutions: Combined effects of correlations and excluded volume. *The Journal of Chemical Physics* **2018,** *149* (5), 054504.
40. Pashley, R. M., Forces between mica surfaces in $La^{3+}$ and $Cr^{3+}$ electrolyte-solutions. *Journal of Colloid and Interface Science* **1984,** *102* (1), 23-35.
41. Agrawal, N. R.; Wang, R., Electrostatic correlation induced ion condensation and charge inversion in multivalent electrolytes. *Journal of Chemical Theory and Computation* **2022,** *18*, 6271-6280.
42. Prashant Sinha; Szilagyi, I.; Ruiz-Cabello, F. J. M.; Maroni, P.; Borkovec, M., Attractive forces between charged

colloidal particles induced by multivalent ions revealed by confronting aggregation and direct force measurements. *The Journal of Physical Chemistry Letters* **2013,** *4*, 648.
43. Bjerrum, N. K., Math-fys. Medd. *Dan. Vidensk. Selsk.* **1926,** *7* (9), 1.
44. Fuoss, R. M., Review of the Theory of Electrolytic Conductance. *Journal of Solution Chemistry* **1978,** *7* (10), 771.
45. Marcus, Y.; Hefter, G., Ion Pairing. *Chemical Reviews* **2006,** *106*, 4585-4621.
46. Adar, R. M.; Markovich, T.; Andelman, D., Bjerrum pairs in ionic solutions: A Poisson-Boltzmann approach. *The Journal of Chemical Physics* **2017,** *146*, 194904.
47. Sessa, F.; Busato, M.; Meneghini, C.; Migliorati, V.; Lapi, A.; D'Angelo, P., Evolution of the La(III) ion coordination sphere in ethylammonium nitrate solution upon water addition. *Journal of Molecular Liquids* **2023,** *388*, 122771.
48. Agrawal, N. R.; Duan, C.; Wang, R., Nature of overcharging and charge inversion in electrical double layers. *The Journal of Physical Chemistry B* **2023,** *128* (1), 303-311.
49. Israelachvili, J. N., Intermolecular and Surface Forces, 3rd Edition. *Intermolecular and Surface Forces, 3rd Edition* **2011**, 1-674.
50. Cygan, R. T.; Liang, J. J.; Kalinichev, A. G., Molecular models of hydroxide, oxyhydroxide, and clay phases and the development of a general force field. *J. Phys. Chem. B* **2004,** *108* (4), 1255-1266.
51. Berendsen, H. J. C.; Grigera, J. R.; Straatsma, T. P., The missing term in effective pair potentials. *J. Phys. Chem.* **1987,** *91* (24), 6269-6271.
52. Migliorati, V.; Serva, A.; Terenzio, F. M.; D'Angelo, P., Development of Lennard-Jones and Buckingham Potentials for Lanthanoid Ions in Water. *Inorg Chem* **2017,** *56* (11), 6214-6224.
53. Mosallanejad, S.; Oluwoye, I.; Altarawneh, M.; Gore, J.; Dlugogorski, B. Z., Interfacial and bulk properties of concentrated solutions of ammonium nitrate. *Phys. Chem. Chem. Phys.* **2020,** *22* (47), 27698-27712.
54. Plimpton, S., FAST PARALLEL ALGORITHMS FOR SHORT-RANGE MOLECULAR-DYNAMICS. *J. Comput. Phys.* **1995,** *117* (1), 1-19.
55. Toxvaerd, S.; Dyre, J. C., Communication: Shifted forces in molecular dynamics. *Journal of Chemical Physics* **2011,** *134* (8), 081102.